\documentclass[a4paper,11pt]{amsart}
\usepackage{amssymb}
\begin{document}

\hyphenation{gra-vi-ta-tio-nal re-la-ti-vi-ty Gaus-sian
re-fe-ren-ce re-la-ti-ve gra-vi-ta-tion Schwarz-schild
ac-cor-dingly gra-vi-ta-tio-nal-ly re-la-ti-vi-stic pro-du-cing
de-ri-va-ti-ve ge-ne-ral ex-pli-citly des-cri-bed ma-the-ma-ti-cal
de-si-gnan-do-si coe-ren-za pro-blem gra-vi-ta-ting geo-de-sic
per-ga-mon cos-mo-lo-gi-cal gra-vity cor-res-pon-ding
de-fi-ni-tion phy-si-ka-li-schen ma-the-ma-ti-sches ge-ra-de
Sze-keres con-si-de-red tra-vel-ling ma-ni-fold re-fe-ren-ces
geo-me-tri-cal in-su-pe-rable sup-po-sedly at-tri-bu-table}

\title[About the Lense-Thirring and Thirring effects]
{{\bf About the Lense-Thirring\\and Thirring effects}}

\author[Angelo Loinger]{Angelo Loinger}
\address{A.L. -- Dipartimento di Fisica, Universit\`a di Milano, Via
Celoria, 16 - 20133 Milano (Italy)}
\author[Tiziana Marsico]{Tiziana Marsico}
\address{T.M. -- Liceo Classico ``G. Berchet'', Via della Commenda, 26 - 20122 Milano (Italy)}
\email{angelo.loinger@mi.infn.it} \email{martiz64@libero.it}

\vskip0.50cm

\begin{abstract}
In Part I we prove that in the linearized version of GR the
``dust-like'' model of a spinning sphere cannot give the
Lense-Thirring and Thirring effects. In Part II we give a proper
and model-independent deduction of both effects within the
linearized version of GR.
\end{abstract}

\maketitle


\normalsize

\vskip1.20cm \noindent \textbf{Introduction} -- As it is known, in
recent years various experiments with the object to verify the
real existence of the Lense-Thirring effect \cite{2x} have been
performed. And the researchers have concluded that this effect
exists \cite{5x}, \cite{6x}.

\par The present paper regards exclusively the theoretical aspect
of the question. In Part I we prove that the computations of Lense
and Thirring \cite{2x} have a fundamental defect, which makes
problematic their result. Actually, the \emph{linearized} version
of GR does \emph{not} give any frame-dragging effect \emph{\`a la}
Lense-Thirring, if we employ the ``dust-like'' model for a
spinning sphere. An analogous negative conclusion holds for
Thirring's articles quoted in \cite{3x}. In Part II we deduce both
effects within the linear approximation of GR. Our results are
model-independent and physically adequate -- they have no
components of centrifugal-like accelerations along the rotation
axis of the spheres.

\vskip0.50cm
\begin{center}
\noindent \small PART I
\end{center} \normalsize

\vskip0.50cm \noindent \textbf{1.} -- The linear approximation of
GR allows to start from eqs. (\ref{eq:one}) of Lense-Thirring
\cite{2x}; $\delta_{\mu\nu}$ is the Minkowskian Kronecker tensor:

\begin{displaymath} \label{eq:one}
\left\{ \begin{array}{l} g_{\mu\nu} =  - \delta_{\mu\nu} +
\gamma_{\mu\nu} \quad; \quad  \gamma_{\mu\nu}= \gamma'_{\mu\nu} - \frac{1}{2} \, \delta_{\mu\nu} \, \gamma'_{\alpha\alpha}\quad; \\ \\
\gamma'_{\mu\nu}  =  - \displaystyle \frac{\kappa}{2\pi} \, \int
\frac{T_{\mu\nu}(x', y', z', t-R)}{R} \,\, \textrm{d}V \, ; \quad
(\alpha, \mu, \nu=1,2,3,4 \,; \quad x_{4}=it): \tag{1}
\end{array} \right.
\end{displaymath}

$\kappa$ is the Einsteinian gravitational constant;
$\textrm{d}V=r'^{2}\textrm{d}r' \sin\vartheta' \,
\textrm{d}\vartheta' \, \textrm{d}\varphi'$; $R$ is the distance
between the field point and the integration element. For
$T_{\mu\nu}$ the Authors choose (for simplicity) the energy tensor
of a ``cloud of dust'', \emph{i.e.}:

\setcounter{equation}{1}
\begin{equation} \label{eq:two}
T_{\mu\nu} = T^{\mu\nu} = \varrho_{0} \,
\frac{\textrm{d}x_{\mu}}{\textrm{d}s} \,
\frac{\textrm{d}x_{\nu}}{\textrm{d}s} \quad ;
\end{equation}

the scalar $\varrho_{0}$ is the invariant mass density. The
\emph{postulated} velocity components are:

\begin{displaymath} \label{eq:three}
\left\{ \begin{array}{l} \displaystyle
\frac{\textrm{d}x_{1}}{\textrm{d}x_{4}} = \, -i \,
\frac{\textrm{d}x'}{\textrm{d}t} = i \, r' \, \omega \,
\sin\vartheta'
 \, \sin\varphi' \quad, \\ \\
\displaystyle \frac{\textrm{d}x_{2}}{\textrm{d}x_{4}} = \, -i \,
\frac{\textrm{d}y'}{\textrm{d}t} = -i \, r' \, \omega \,
\sin\vartheta'
 \, \cos\varphi' \quad, \\ \\
\displaystyle \frac{\textrm{d}x_{3}}{\textrm{d}x_{4}} = \, 0
\quad; \tag{3}
\end{array} \right.
\end{displaymath}

$r'$, $\vartheta'$, $\varphi'$ are the polar coordinates of a
point of the rotating sphere of ``dust''; the rotation happens
round the $Z$-axis with angular velocity $\omega$.

\par Formulae (\ref{eq:three}) play an essential role in the
computations of the papers \cite{2x}, \cite{3x}. Unfortunately,
these formulae are \emph{not} consistent with the equations of
motion of the ``dust'' particles that are \emph{prescribed} by the
linear theory, as we shall prove.

\vskip1.20cm \noindent \textbf{2.} -- It is well known that in the
linear approximation of GR the \emph{ordinary} divergence of
$T_{\mu\nu}$ is equal to zero. Now, in 1918 people did not know
that the equations of matter motion are an analytical consequence
of the gravitational field equations (\textbf{\emph{both}}
\emph{in the exact GR} \textbf{\emph{and}} \emph{in its linear
approximation} \cite{2}). In our case, the equations $\partial
T_{\mu\nu} / \partial x_{\nu} =0$ imply the equations of motion of
the ``dust'' particles. But these equations tell us that the
particles describe \emph{rectilinear and uniform} motions;
\emph{any acceleration is excluded}, in particular any rotation.
The proof is quite trivial; let us consider the equations

\setcounter{equation}{3}
\begin{equation} \label{eq:four}
\frac{\partial}{\partial x_{\nu}}\left( \varrho_{0} \, u_{\mu} \,
u_{\nu} \right) = 0 \quad,
\end{equation}

where $u_{\mu}=\textrm{d}x_{\mu} /\textrm{d}s$; from which

\begin{equation} \label{eq:five}
u_{\mu} \, \frac{\partial}{\partial x_{\nu}} (\varrho_{0}u_{\nu})
 + \varrho_{0} \, u_{\nu} \, \frac{\partial u_{\mu}}{\partial
x_{\nu}}= 0 \quad;
\end{equation}

multiply these equations by $u_{\mu}$; the second term gives zero
(because $u_{\mu} \, \partial u_{\mu} / \partial x_{\nu}=0$; we
are left with $\partial (\varrho_{0}u_{\nu})/\partial x_{\nu}=0$,
the conservation equation of matter -- and eqs. (\ref{eq:five})
reduce to

\begin{equation} \label{eq:six}
u_{\nu} \, \frac{\partial u_{\mu}}{\partial x_{\nu}} = 0 \quad,
\end{equation}

\emph{i.e.}:

\begin{equation} \label{eq:seven}
\frac{\textrm{d}u_{\mu}}{\textrm{d}s} = 0 \quad, \quad
\textrm{\emph{q.e.d.}}
\end{equation}

This means that Lense-Thirring and Thirring effects could be
\emph{possibly} obtained only with computations that go
\emph{beyond the linear stage} of \emph{approximation}, if we
adopt the ``dust-like'' model for the spinning spheres. Such
conclusion holds also for the frame-dragging effect induced by a
generic non-zero four-acceleration of the ``dust'' particles
\cite{4x}.

\vskip1.20cm \noindent \textbf{3.} -- In their papers \cite{2x},
\cite{3x} Thirring and Lense employed, \emph{more antiquo}, an
imaginary time coordinate $x_{4}=it$. This makes particularly
intuitive at any computational step that the considered spacetime
$\mathfrak{P}$ is pseudo-Euclidean, \emph{i.e.} that we are
dealing with the special relativity. The $\gamma_{\mu\nu}$'s are
the components of a symmetrical tensor-field of the second order
in the Minkowskian spacetime $\mathfrak{P}$.

\par The field equations of the linear approximation of GR remain
unchanged if $\gamma_{\mu\nu}$ is replaced by

\begin{equation} \label{eq:A1}
\gamma'_{\mu\nu} = \gamma_{\mu\nu} + \frac{\partial \, \xi_{\mu}
}{\partial \, x_{\nu}} + \frac{\partial \, \xi_{\nu} }{\partial \,
x_{\mu}} \quad;
\end{equation}

the four functions $\xi_{\mu}$'s can be chosen in such a way that
$\gamma'_{\mu\nu} = \gamma_{\mu\nu} -(1/2)\, \delta_{\mu\nu}\,
\gamma_{\alpha\alpha}$ satisfies the equations

\begin{equation} \label{eq:A2}
\frac{\partial \, \gamma'_{\mu\nu} }{\partial \, x_{\nu}}=0 \quad,
\end{equation}

in correspondence with the differential conservation equations
$\partial T_{\mu\nu}/ \partial x_{\nu}=0$.

\par Eqs. (\ref{eq:A1}) have a \emph{twofold} interpretation:
\emph{i}) from the Minkowskian standpoint they represent a gauge
transformation which is quite analogous to a gauge transformation
of Maxwell electrodynamics \cite{2}; \emph{ii}) from a general
standpoint they are the result of an infinitesimal coordinate
transformation $x'_{\mu}=x_{\mu}+ \xi_{\mu}(x)$ on
$g_{\mu\nu}=-\delta_{\mu\nu}+ \gamma_{\mu\nu}$.

\par A last remark. Of course, the physically legitimate
$\gamma_{\mu\nu}$'s allow to compute the geodesic of a
\emph{test}-particle in the gravitational potential
$g_{\mu\nu}=-\delta_{\mu\nu}+ \gamma_{\mu\nu}$.

\vskip0.50cm
\begin{center}
\noindent \small PART II
\end{center} \normalsize

\vskip0.50cm \noindent \textbf{4.} -- \emph{\textbf{Frame-dragging
by a spinning full sphere}} -- We use the previous notations, but
employing the \emph{CGS system of units}. The fundamental
equations of the linear approxiamtion of GR are:

\begin{equation} \label{eq:onex}
\Box \, \gamma'_{\mu\nu}  =  \displaystyle \frac{\kappa}{2\pi} \,
T_{\mu\nu} \quad ; \quad (x_{4}=ict) \quad.
\end{equation}

It follows from eqs.(\ref{eq:onex}) that any material homogeneous
sphere $\Sigma$ (mass $M$, radius $l$), which is at rest in our
Cartesian orthogonal system $S(X_{1},X_{2},X_{3})$, with its
centre at $X_{1}=X_{2}=X_{3}=0$, generates an external
gravitational potential given by:

\setcounter{equation}{10}
\begin{equation} \label{eq:twox}
\gamma'_{44}{}^{(0)} = \frac{4GM/c^{2}}{R} \equiv \frac{4m}{R}
\quad ; \quad \gamma'_{\mu\nu}{}^{(0)} = 0 \quad, \quad
\textrm{if} \quad (\mu,\nu) \neq (4,4) \quad ,
\end{equation}

where $R=(X_{1}^{2}+X_{2}^{2}+X_{3}^{2})^{1/2} \geq l$ .

Let us now assume that $\Sigma$ rotates round the $X_{3}$-axis,
with a small angular velocity $\omega$, with respect to $S$. We
compute the new $\gamma'_{\mu\nu}$'s, neglecting with Lense and
Thirring (and for the identical reasons) the $\gamma'_{\mu\nu}$'s
which contain $\omega^{2}$. Then, we compute the corresponding
$\gamma_{\mu\nu}$'s. We get, if $\chi \equiv \omega / c$:

\begin{displaymath} \label{eq:threex}
\left\{ \begin{array}{l} \displaystyle \gamma_{11} = - \frac{2m}R
\quad; \quad  \gamma_{12} \approx 0 \quad ; \quad  \gamma_{13} = 0 \quad; \quad  \gamma_{14}= \frac{4m i \chi}{R}\, X_{2} \quad ;\\ \\
\displaystyle \gamma_{22} = - \frac{2m}R
\quad; \quad  \gamma_{23} = 0 \quad ; \quad  \gamma_{24}= -\frac{4m i \chi}{R}\, X_{1} \quad ; \\ \\
\displaystyle \gamma_{33} = - \frac{2m}R \quad; \quad  \gamma_{34}
= 0 \quad ; \quad  \gamma_{44}= - \frac{2m}R \quad . \tag{12}
\end{array} \right.
\end{displaymath}

Denoting with the small letters $x_{1},x_{2},x_{3},x_{4}, r$ the
time-dependent dynamical variables which concern the geodesic
motion of a test-particle through the gravitational field of
eqs.(\ref{eq:threex}), we have $(j=1,2,3)$:

\setcounter{equation}{12}
\begin{equation} \label{eq:fourx}
\ddot{x}_{j}= \frac{1}{2} \left(\gamma_{j\mu,\nu} +
\gamma_{j\nu,\mu}- \gamma_{\mu\nu,j}\right) \, \dot{x}_{\mu} \,
\dot{x}_{\nu} \quad,
\end{equation}

where an overdot means a derivative with respect to time $t$, and
the comma the derivative with respect to a coordinate
$x_{\lambda}$. Eqs.(\ref{eq:fourx}) are the equations of a
geodesic with the approximation $ic\textrm{d}t \approx
\textrm{d}s$ -- the squares of all three-velocity components
divided by $c$ are neglected (as in \cite{2x}). And in the
three-acceleration $\ddot{x}_{j}$ we retain only the terms with
indices $(\mu, \nu) = (14), (24), (34), (44)$. We have finally, if
$(x_{1}, x_{2}, x_{3}) \equiv (x,y,z)$:

\begin{displaymath} \label{eq:fivex}
\left\{ \begin{array}{l} \displaystyle \ddot{x} = 4m\omega \left(
\frac{x^{2}+y^{2}}{r^{3}} - \frac{3}{r}\right) \dot{y} +  4m\omega
y \left(\frac{x\dot{x}+y\dot{y}+2z\dot{z}}{r^{3}}\right) -
\frac{GM}{r^{2}} \, \frac{x}{r}\quad
; \\ \\
\displaystyle \ddot{y} = -4m\omega \left(
\frac{x^{2}+y^{2}}{r^{3}} - \frac{3}{r}\right) \dot{x} -  4m\omega
x
\left(\frac{x\dot{x}+y\dot{y}+2z\dot{z}}{r^{3}}\right) - \frac{GM}{r^{2}} \, \frac{y}{r}\quad ; \\ \\
\displaystyle \ddot{z} = - \frac{GM}{r^{2}} \, \frac{z}{r} \quad.
\tag{14}
\end{array} \right.
\end{displaymath}

In Appendix A we have transcribed formulae $(15)$ of Lense and
Thirring \cite{2x} in the CGS system of units. The z-component of
their acceleration contains the Coriolis-like term
$(m/r^{2})(\omega l^{2}/r)[(12z/5r)\cdot(x\dot{y}-y\dot{x})/r]$.
\emph{This would be sufficient to conclude that their results are
problematic.}

\vskip1.20cm \noindent \textbf{5.} -- \emph{\textbf{Frame-dragging
by a spinning hollow sphere}} -- We consider with Thirring an
infinitely thin spherical shell $\Sigma$ of radius $a$ and mass
$M$. If this shell is at rest in $S$, the \emph{internal}
gravitational potential can be suitably written as follows:

\setcounter{equation}{14}
\begin{equation} \label{eq:sixx}
\gamma'_{44}{}^{(0)} = \frac{4m}{a} \quad ; \quad
\gamma'_{\mu\nu}{}^{(0)} = 0 \quad, \quad \textrm{if} \quad
(\mu,\nu) \neq (4,4) \quad ;
\end{equation}

this expression of the (constant) internal potential coincides
with the value for $R=a$ of the external potential $4m/R$. Assume
now that $\Sigma$ rotates round the $X_{3}$-axis. With Thirring
\cite{3x}, in the computation of the $\gamma'_{\mu\nu}$'s we take
into consideration also the terms with $\omega^{2}(\equiv
c^{2}\chi^{2})$, that will give a kind of centrifugal force.

\par Derive the $\gamma_{\mu\nu}$'s from the
$\gamma'_{\mu\nu}$'s; putting for brevity $K\equiv 4m/a$, we have:

\begin{displaymath} \label{eq:sevenx}
\left\{ \begin{array}{l} \displaystyle \gamma_{11} = \frac{1}{2}
\, K
-  \frac{1}{2} \, K \chi^{2} \left(X_{2}^{2}-X_{1}^{2}\right) \quad; \quad  \gamma_{12} = K \chi^{2}X_{1}X_{2} \quad; \\ \\
\displaystyle \gamma_{13} = 0 \quad; \quad  \gamma_{14} = K i \chi
X_{2} \quad; \tag{16}
\end{array} \right.
\end{displaymath}

\begin{displaymath} \label{eq:sevenprimex}
\left\{ \begin{array}{l} \displaystyle \gamma_{22} = - \frac{1}{2}
\, K
-  \frac{1}{2} \, K \chi^{2} \left(X_{1}^{2}-X_{2}^{2}\right) \quad; \quad  \gamma_{23} = 0  \quad; \\ \\
\displaystyle \gamma_{24} = - K i \chi X_{1} \quad; \tag{16$'$}
\end{array} \right.
\end{displaymath}

\begin{displaymath} \label{eq:sevensecondx}
\left\{ \begin{array}{l} \displaystyle \gamma_{33} = - \frac{1}{2}
\, K
+  \frac{1}{2} \, K \chi^{2} \left(X_{1}^{2}+X_{2}^{2}\right) \quad; \quad  \gamma_{34} = 0 \quad; \\ \\
\displaystyle \gamma_{44} = \frac{1}{2} \, K +  \frac{1}{2} \, K
\chi^{2} \left(X_{1}^{2}+X_{2}^{2}\right) \quad. \tag{16$''$}
\end{array} \right.
\end{displaymath}

The equations of the geodesic of a test-particle in the
gravitational field of these $\gamma_{\mu\nu}$'s are identical to
eqs.(\ref{eq:fourx}). And, as in sect. \textbf{4}, we consider
only the terms with indices $(\mu, \nu) = (14), (24), (34), (44)$.
We get:

\begin{displaymath} \label{eq:eigthx}
\left\{ \begin{array}{l} \displaystyle \ddot{x} = -12 \,
\frac{m}{a} \, \omega \, \dot{y} + 2 \, \frac{m}{a} \,
\omega^{2}\, x  \quad; \\ \\
\displaystyle \ddot{y} = 12 \, \frac{m}{a} \, \omega \, \dot{x} +
2 \, \frac{m}{a} \,
\omega^{2}\,y  \quad; \\ \\
\ddot{z} = 0 \quad. \tag{17}
\end{array} \right.
\end{displaymath}

In Appendix B we have transcribed formulae $(22)$ of Thirring
paper \cite{3x} in the CGS system of units. \emph{Thirring's
formula for $\ddot{z}$ is a clear absurdity from the physical
standpoint.} He tried to justify it with the following sentences:
``Die dritte Gleichung $[(22)]$ liefert das im ersten Augenblick
\"uberraschende Ergebnis, da\ss {} diese ``Zentrifugalkraft'' noch
eine axiale Komponente besitzt $[$\emph{i.e.},
$-(8m/15a)\,\omega^{2}z]$. Ihr Auftreten im Felde der rotierenden
Kugel l\"a\ss t sich folgenderma\ss en aufkl\"aren: Vom ruhenden
Beobachter aus betrachtet haben jene Fl\"achenelemente der
Hohlkugel, welche sich in der N\"ahe des \"Aquators befinden,
gro\ss ere Geschwindigkeit, und infolgedessen auch gr\"o\ss ere
scheinbare (tr\"age und gravitierende) Masse als jene, die sich in
der Umgebung der Pole befinden. Das Feld einer mit konstanter
Fl\"achendichte belegten rotierenden Hohlkugel entspricht also dem
einer ruhenden Kugelschale, bei welcher die Fl\"achendichte mit
wachsendem Polabstand $\vartheta$ zunimmt. Da\ss {} im letzteren
Falle Punkte $[$\emph{i.e.} test-particles$]$, die au\ss erhalb
der \"Aquatorebene liegen, in sie hineingezogen werden, ist ohne
weiteres verst\"andlich.'' -- This is a clever, but illogical (and
\emph{a posteriori}) justification. Indeed, in the deduction of
Thirring's formulae $(22)$ there is no use of a possible
difference (by virtue of the difference of the respective
velocities) between the masses of the equatorial zone and the
masses of the polar zones.

\vskip0.50cm \par We hope that our deduction of the Lense-Thirring
effect will be appreciated by the
 teams of experimentalists, who, after years of refined efforts,
 have concluded in favour of the existence of this effect -- notwithstanding
 the subtle difficulties to take properly
 into account the systematic errors of measurement. --

\vskip2.00cm
\begin{center}
\noindent \small \emph{\textbf{APPENDIX A}}
\end{center} \normalsize

\vskip0.40cm \noindent Formulae $(15)$ of Lense and Thirring
\cite{2x} in the CGS system of units ($m\equiv GM/c^{2}$; $l$ is
the radius of the sphere):

\begin{displaymath} \label{eq:A1x}
\left\{ \begin{array}{l} \displaystyle \displaystyle \ddot{x} =
\frac{m}{r^{2}} \, \frac{\omega l^{2}}{r} \, \left[\frac{4}{5}\,
\frac{x^{2}+y^{2}-2z^{2}}{r^{2}}\, \dot{y} + \frac{12}{5} \,
\frac{yz}{r^{2}} \, \dot{z}\right]
 - \frac{GM}{r^{2}} \, \frac{x}{r}\quad; \\ \\
\displaystyle \ddot{y} = - \frac{m}{r^{2}} \, \frac{\omega
l^{2}}{r} \, \left[\frac{4}{5}\,
\frac{x^{2}+y^{2}-2z^{2}}{r^{2}}\, \dot{x} + \frac{12}{5} \,
\frac{xz}{r^{2}} \, \dot{z}\right]
 - \frac{GM}{r^{2}} \, \frac{y}{r}\quad; \\ \\
\displaystyle \ddot{z} = \frac{m}{r^{2}} \, \frac{\omega l^{2}}{r}
\, \left[\frac{12}{5}\, \frac{z}{r} \,
\frac{x\dot{y}-y\dot{x}}{r}\right]
 - \frac{GM}{r^{2}} \, \frac{z}{r}\quad.
\end{array} \right.
\end{displaymath}

\vskip2.00cm
\begin{center}
\noindent \small \emph{\textbf{APPENDIX B}}
\end{center} \normalsize

Formulae $(22)$ of Thirring \cite{3x} in the CGS system of units
($m\equiv GM/c^{2}$; $a$ is the radius of the spherical shell):

\begin{displaymath} \label{eq:B1x}
\left\{ \begin{array}{l} \displaystyle \ddot{x} = - \frac{8m}{3a}
\, \omega \, \dot{y} + \frac{4}{15} \, \frac{m}{a} \,
\omega^{2}\, x  \quad; \\ \\
\displaystyle \ddot{y} = \frac{8m}{3a} \, \omega \, \dot{x} +
\frac{4}{15} \, \frac{m}{a} \,
\omega^{2}\, y   \quad; \\ \\
\displaystyle \ddot{z} = - \frac{8m}{15a} \, \omega^{2} \, z
\quad.
\end{array} \right.
\end{displaymath}

(We have taken into account the \emph{Berichtigung} of 1921, see
\cite{3x} \emph{ii}). --

\end{document}